\documentclass[a4paper,12pt,openany]{article}
\usepackage[top=20truemm,bottom=20truemm,left=23truemm,right=23truemm]{geometry}
\usepackage[]{hyperref}
\usepackage{amsmath}
\usepackage{ascmac}
\usepackage{amssymb}
\usepackage{ulem}
\usepackage[]{graphicx}
\usepackage[utf8]{inputenc}

\graphicspath{{./figs/}}

\title{Time dependent Interface in AdS Black Hole Spacetime}
\author{Koichi Nagasaki}
\date{\today}

\begin{document}
\vspace{1cm}

\begin{center}
	{\LARGE Time dependent Interface in AdS Black Hole Spacetime}\\
\vspace{2cm}
	{\large Koichi Nagasaki}\footnote{koichi.nagasaki24@gmail.com}\\
\vspace{1cm}
	{\small School of Physics, 
	University of Electronic Science and Technology of China,\\
	Address: No.4, Section 2, North Jianshe Road, Chengdu, 
	Sichuan 610054, China}
\end{center}
\vspace{1.5cm}

\abstract{We consider a D5-brane solution in AdS black hole spacetime.
This is a defect solution moving in subspace of AdS$_5\times S^5$.
This non-local object is realized by the probe D5-brane moving in black hole spacetime.
We found this probe brane does not penetrate the black hole horizon.
We also found the solution does not depend on the motion on $S^5$ subspace.
}
\tableofcontents

\section{Introduction}
In the AdS/CFT correspondence \cite{Maldacena:1997re, Witten:1998qj, Gubser:1998bc} non-local operators are a useful tool for studying. 
For example, in \cite{Nagasaki:2011ue}, we considered a kind of non-local operator called "an interface" and found the agreement in the results of gauge theory and gravity theory in the leading order of the power series of $\lambda/k^2$.
Here $\lambda$ is the 't Hooft coupling $\lambda = g_\text{YM}^2N$ and $k$ is a parameter which specifies the gauge flux on the probe brane.
This non-local operator is realized in the D3/D5 brane system.
The multiple D3-brane create a curved space-time and one D5-brane is treated as a probe on the AdS$_5\times S^5$ spacetime.
These brane systems are related to nonequilibrium systems in the AdS/CFT correspondence \cite{Karch:2002sh, Constable:2002xt, Evans:2010hi, Das:2010yw}.
A time dependent solution of the D5-brane we consider in this paper has its application to cosmology in anti de-Sitter space and brane world scenario \cite{Banerjee:2018qey, Banerjee:2019fzz}. 

\begin{figure}[h]
\begin{center}
	\includegraphics[width=\linewidth]{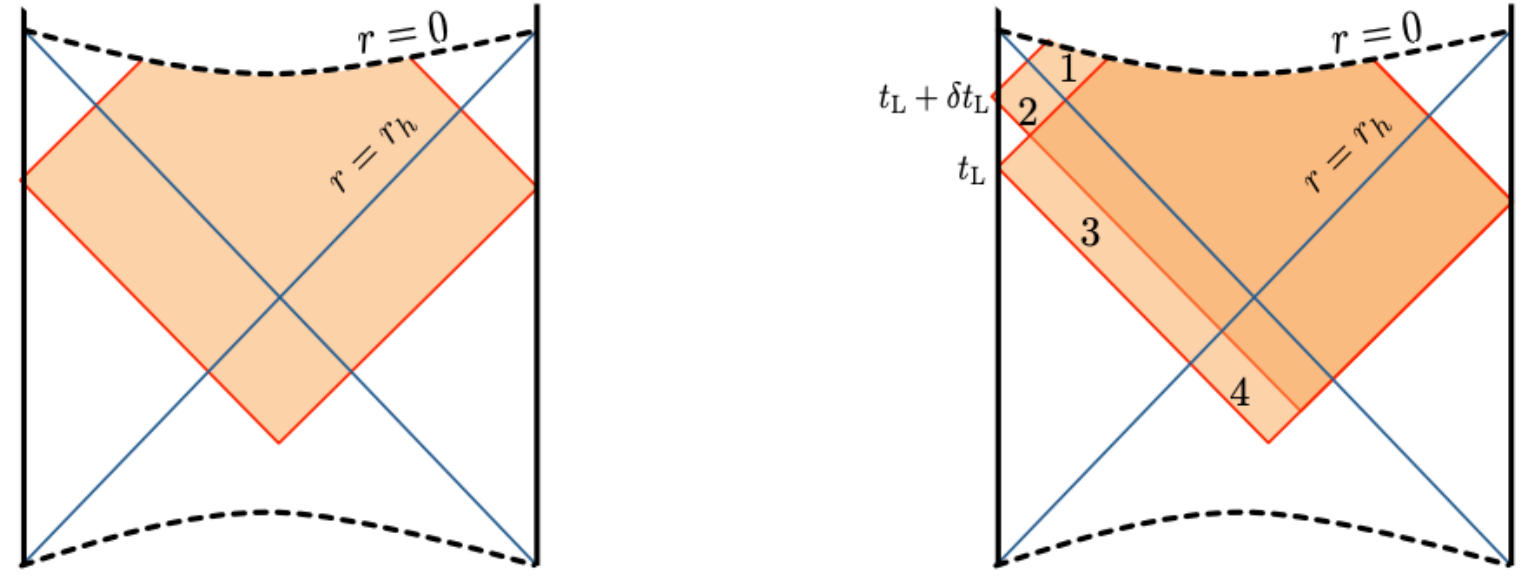}
	\caption{Left: Wheeler DeWitt patch. Right: Development of the region}
	\label{fig:AdSBH_WDW}
\end{center}
\end{figure}

The motivation to consider the AdS/CFT correspondence including the D5-brane on black hole spacetimes is related to a conjecture called ``complexity - action" (CA).
In this statement complexity is a quantity which is expected to be related many black hole problems \cite{Susskind:2012rm, Almheiri:2012rt, Susskind:2014moa, Susskind:2014rva}.
Complexity has the origin in computational science and it quantifies how hard to create the final state from the initial state \cite{0034-4885-75-2-022001,2008arXiv0804.3401W,Arora:2009:CCM:1540612,Moore:2011:NC:2086753,Susskind:2014moa,Susskind:2014rva,Caputa:2017yrh, Hashimoto:2018bmb, Bhattacharyya:2018wym}.
In black hole physics this quantity describes a quantum state of the black hole interior or Einstein-Rosen Bridge \cite{PhysRev.48.73}.
According to the holographic relation stated in the first paragraph, complexity must have the holographic counterpart. 
That is called holographic complexity. 
CA conjecture \cite{Brown:2015bva, Brown:2015lvg} states that this holographic quantity is the action which is calculated in a bulk region called ``the Wheeler DeWitt patch."
This is the region bounded by null surfaces anchored at the given time on boundaries.
The calculation of the action has a difficulty caused by the divergence at the AdS boundary where the metric diverges.
To compute the growth rate of the action, we only need to take into account the bulk region inside the horizon as explained in \cite{Nagasaki:2017kqe}.
The left panel of Figure \ref{fig:AdSBH_WDW} depicts Wheeler DeWitt patch on a Penrose diagram.
The two diagonal lines represent the black hole horizon.
As the time passes, this region changes as shown in the right of Figure \ref{fig:AdSBH_WDW}.
The time defined on a boundary theory is denoted by $t_L$.
The separated regions (2), (3) and (4) do not contribute the development.
We only have to find the contribution of region (1) to measure the time development of the WDW action.
   
We found the static defect solutions in the AdS black hole spacetime in \cite{Nagasaki:2018dyz, Nagasaki:2019fln}.
These describes the embedding of the D5-brane in AdS black hole spacetime.
The results tell us that the D5-brane can only extend outside of the horizon if there is a non-zero gauge flux on the D5-brane.
In CA conjecture holographic complexity --- the action is calculated by the integral over the Wheeler-DeWitt patch. 
Especially, in order to find the growth rate of it, we only have to integrate inside the horizon. 
Then the DBI action does not contribute the growth of complexity in these cases.

A time dependent system is interesting for studying the AdS/CFT correspondence.
Its behavior is different from the static cases.
For example, the extremal surfaces do not penetrate the horizon in a static case, whereas it can penetrate the horizon in the time dependent case as stated in \cite{Hubeny:2012ry}.
Then it will be worth to consider a moving defect \cite{Janiszewski:2011ue} in AdS black hole spacetime.
These time dependent cases have never been studied for the AdS/CFT correspondence and CA conjecture. 
If such a solution is found, this can be a good way to test CA relation.
As stated in the first, a new parameter is useful to compare the holographic quantities.

In the above reason we would like to consider a moving object in the AdS black hole spacetime.
The whole spacetime consists of the product of AdS$_5$ and $S^5$.
We can suppose some types of motion of the D-branes.
For simplicity, in this paper we would like to consider a motion which has a spherical symmetry which has the rotation in $S^2$ subspace of $S^5$. 

The organization of this paper is as follows.
In Section \ref{sec:flatspacetime} we start with the study of the flat AdS case and find its solution.
This case corresponds to the flat spacetime solution \cite{Nagasaki:2011ue} added the angular momentum.
In Section \ref{sec:BHspacetime} we generalize the solution to the black hole spacetime.
We conclude this paper in Section \ref{sec:Discussion} with some comments for the results.

\section{Flat spacetime}\label{sec:flatspacetime}
In this section we consider the embedding of the D5-brane in AdS spacetime AdS$_5\times S^5$.
The metric of the AdS part with radius $R_\text{AdS}$ is given by time coordinate $t$, cartesian coordinates $x_i; i= 1,2,3$ and the radial coordinate $1/y$ as
\begin{equation}
ds_\text{AdS}^2 = \frac{R_\text{AdS}^2}{y^2}(-dt^2+dy^2+\sum_{i=1}^3dx_i^2).
\end{equation}
The $S^5$ metric is given by coordinates $\theta\in[0,\pi/2]$, $\varphi\in[0,2\pi)$, 
$\psi\in[0,\pi/2]$ and $\phi,\chi\in[0,2\pi)$ as
\begin{align}
ds_{S^5}^2
&= d\theta^2 + \sin^2\theta d\varphi^2 + \cos^2\theta d\Omega_3^2\nonumber\\
&= d\theta^2 + \sin^2\theta d\varphi^2 + \cos^2\theta
  (d\psi^2 + \sin^2\psi d\phi^2 + \cos^2\psi d\chi^2).
\end{align}
In the following we put $R_\text{AdS} = 1$ for simplicity.
The Ramond-Ramond 4-form is defined as
\begin{equation}\label{eq:TiAdSfl_RRform}
C_4 = -\frac1{y^4}dtdx_1dx_2dx_3 + 4\alpha_4,
\end{equation}
where $\alpha_4$ is a 4-form satisfying $d\alpha_4 = \text{volume of }S^5$.

We consider the D5-brane rotates in subspace of $S^5$.
We assume that the D5-brane extends to the directions
\begin{equation}
t,x_1,x_2\in\text{AdS}_5,\;
\psi,\phi\in S^5,
\end{equation}
and one dimensional subspace in other directions. 
By parameter $\sigma$, this embedding is given by the following functions
\begin{equation}\label{eq:TiAdSfl_directions}
y = y(\sigma),\;\;
x_3 = x_3(\sigma),\;\;
\theta = \theta(\sigma),\;\;
\varphi = \omega t + g(\sigma).
\end{equation}
There is a gauge flux on the D5-brane which is expressed by symmetry as
\begin{equation}\label{eq:TiAdSfl_gaugeflux1}
\mathcal F = F'(\psi)d\psi\wedge d\phi,\;\;
\mathcal A = F(\psi)d\phi.
\end{equation}
In this section ${}^\prime$ means the derivative by the variables indicated in eq.\eqref{eq:TiAdSfl_directions} and eq.\eqref{eq:TiAdSfl_gaugeflux1}.
The induced metric is
\begin{align*}
ds_\text{D5}^2
&= -\Big(\frac1{y^2} - \omega^2\sin^2\theta\Big)dt^2
 + \Big(\frac{y'^2+x_3'^2}{y^2} + \theta'^2 + g'^2\sin^2\theta\Big)d\sigma^2 
 + 2\omega g'\sin^2\theta dtd\sigma\nonumber\\
&\hspace{7cm}
 + \frac{dx_1^2+dx_2^2}{y^2}
 + \cos^2\theta(d\psi^2 + \sin^2\psi d\phi^2).
\end{align*}
Adding the flux, the above metric is in matrix form
\begin{align}
&G_\text{ind} + \mathcal F\nonumber\\
&= \begin{bmatrix}
-\frac1{y^2}(1-\omega^2y^2\sin^2\theta)& \omega g'\sin^2\theta& & & & \\
\omega g'\sin^2\theta& \frac1{y^2}(y'^2+x_3'^2+y^2\theta'^2+y^2g'^2\sin^2\theta)& & & & \\
 & & 1/y^2& & & \\
 & & & 1/y^2 & & \\
 & & & & \cos^2\theta& F'\\
 & & & & -F'& \cos^2\theta\sin^2\psi
\end{bmatrix}.
\end{align}
Then the DBI action is 
\begin{equation}
S_\text{DBI}/T_5
= -2\pi{\mathcal T}V\int dyd\psi
  \frac{\sqrt{\cos^4\theta\sin^2\psi + F'^2}}{y^4}L(\sigma),
\end{equation}
where $\mathcal T$ is the integral over the time interval and $V$ is the integral on $(x_1,x_2)$- plane.
The Wess-Zumino action is, by substituting the RR4-form \eqref{eq:TiAdSfl_RRform},
\begin{align}
S_\text{WZ}/T_5 
= \int\mathcal F\wedge C_4
= -2\pi{\mathcal T}V\int\frac{F'x_3'd\psi d\sigma}{y^4}.
\end{align}
Summing them, the D5-brane action is 
\begin{equation}
S_\text{D5}/T_5
= -2\pi{\mathcal T}V\int\frac{d\psi d\sigma}{y^4}\Big(
  \sqrt{\cos^4\theta\sin^2\psi+F'(\psi)^2}L(\sigma) + F'(\psi)x_3'\Big).
\end{equation}

The equation of motion for $F(\psi)$ is 
\begin{equation}
\frac{d}{d\psi}\Big(\frac{F'(\psi)}{\sqrt{\cos^4\theta\sin^2\psi+F(\psi)'^2}}\Big) = 0.
\end{equation}
By using $c(\sigma)$, which is a function of $\sigma$,
\begin{equation}
F'^2 = \frac{c^2\cos^4\theta}{1-c^2}\sin^2\psi,\;
F' = -\kappa\sin\psi,
\end{equation}
where we denote the constant factor in the front of $\sin\psi$ as $\kappa$ since we know $F'$ is independent of $y$:
\begin{equation}
\frac{c(\sigma)^2\cos^4\theta(\sigma)}{1-c(\sigma)^2} =: \kappa^2.
\end{equation}
This constant represents the strength of the gauge flux.
Therefore we found the ansatz for the gauge flux to be proportional to the volume form of $S^2$ subspace:
\begin{equation}\label{eq:TiAdSfl_gaugeflux2}
\mathcal F = -\kappa\sin\psi d\psi\wedge d\phi.
\end{equation}
Summarizing the above discussion, the action is 
\begin{align}
S_\text{D5}/T_5
&= -2\pi TV\int\frac{d\sigma}{y^4}\Big(
  \sqrt{\cos^4\theta+\kappa^2}L(\sigma) - \kappa x_3'
\Big),\\
L(\sigma) &:= \sqrt{(y'^2+x_3'^2+y^2\theta'^2)(1-\omega^2y^2\sin^2\theta)+y^2g'^2\sin^2\theta}.
\end{align}
For simplicity we use the following notation in the next section.
\begin{equation}
\Theta := \sqrt{\cos^4\theta+\kappa^2},\;\;
S := y'^2 + x_3'^2 + y^2\theta'^2,\;\;
\Omega := 1-\omega^2y^2\sin^2\theta.
\end{equation}
By this notation the Lagrangian is 
\begin{equation}
\mathcal L = \frac{\Theta L(\sigma) - \kappa x_3'}{y^4},\;\;
L(\sigma) = \sqrt{S\Omega + y^2g'^2\sin^2\theta}.
\end{equation}

\subsection{Equations of motion}
The equations of motion are
\begin{subequations}\label{eq:TiAdS_4eom}
\begin{align}
&\frac{d}{d\sigma}\Big(\frac{y'\Theta\Omega}{y^4L}\Big)
 + \frac{4}{y^5}(\Theta L - \kappa x_3')
 - \frac{\Theta}{y^3}
  \frac{\theta'^2\Omega + (g'^2-\omega^2S)\sin^2\theta}{L} = 0,\\
&\frac{d}{d\tau}\Big(\theta'
\frac{\Omega\Theta}{y^2L}\Big)
 + \frac{2\sin\theta\cos^3\theta}{y^4\Theta}L
 - \frac{(g'^2-\omega^2S)\sin\theta\cos\theta}{y^2}\frac{\Theta}{L} = 0,\\
&\frac{d}{d\tau}\Big(x_3'\frac{\Omega\Theta}{y^4L}\Big) + \frac{4\kappa y'}{y^5}
= 0,\\
&\frac{d}{d\tau}\Big(g'\frac{\sin^2\theta\cdot\Theta}{y^2L}\Big) = 0.
\end{align}
\end{subequations}
There is a gauge degree of freedom due to reparametrization invariance in $\sigma$. 
We fix $\Theta/L = 1$.
Then the equations \eqref{eq:TiAdS_4eom} are
\begin{subequations}\label{eq:TiAdS_4eom2}
\begin{align}
&\frac{d}{d\sigma}\Big(\frac{y'\Omega}{y^4}\Big)
  + \frac4{y^5}(\Theta^2-\kappa v) 
  - \frac{\theta'^2\Omega + (w^2-\omega^2S)\sin^2\theta}{y^3} = 0,\\
&\frac{d}{d\sigma}\Big(\theta'\frac{\Omega}{y^2}\Big)
 + \frac{2\sin\theta\cos^3\theta}{y^4}
 - \frac{(g'^2-\omega^2S)\sin\theta\cos\theta}{y^2} = 0,\\
&\frac{d}{d\sigma}\Big(x_3'\frac{\Omega}{y^4}\Big) + \frac{4\kappa y'}{y^5} = 0,\\
&\frac{d}{d\sigma}\Big(g'\frac{\sin^2\theta}{y^2}\Big) = 0.
\end{align}
\end{subequations}
We define 
\begin{subequations}
\begin{align}
A &:= -\frac{d}{d\sigma}\log\Omega 
= \frac{2\omega^2y\sin\theta(y'\sin\theta + y\theta'\cos\theta)}{\Omega},\\
B &:= \frac4{y\Omega}(\kappa x_3' - \Theta^2),\;\;
C := \frac{(g'^2-\omega^2S)}\Omega,\;\;
D := -\frac{2\sin\theta\cos^3\theta}{y^2\Omega}.
\end{align}
\end{subequations}
The equations are solved for the second derivative terms as
\begin{subequations}\label{eq:TiAdSfl_4eom}
\begin{align}
y'' &= y\theta'^2 + \frac{4y'^2}y + y'A + B + Cy\sin^2\theta,\\
\theta'' &= \frac{2y'\theta'}y + \theta'A + C\sin\theta\cos\theta + D,\\
x_3'' &= \frac{4y'x_3'}y + x_3'A - \frac{4\kappa y'}{y\Omega},\label{eq:timeintAdSfl_eom_3}\\
g'' &= -2\theta'g'\cot\theta + \frac{2y'g'}y.\label{eq:timeintAdSfl_eom_4}
\end{align}
\end{subequations}

\paragraph{Static case}
In the static case, the solution of the above equations must be $x_3 = \kappa y$ as obtained in \cite{Nagasaki:2011ue}.
Let us confirm it.
For $\omega = 0$ and $\theta = 0$, the equations of motion \eqref{eq:TiAdS_4eom} are simplified by setting $\sigma=y$ as
\begin{equation}
\frac{d}{dy}\Big(
\kappa\frac{\sqrt{1+\kappa^2}}{y^4\sqrt{1+x_3'^2}}\Big)
+ \frac{4\kappa}{y^5}
= 0.
\end{equation}
Then in this case, the following is the solution:
\begin{equation}
x_3 = \kappa y,\;\;
\theta(y) = 0,\;\;
g(y) = 0.
\end{equation}
This is surely the stationary solution \cite{Nagasaki:2011ue}. 

\paragraph{Boundary condition}
$y=0$ corresponds to the AdS boundary.
The boundary condition is, since we fixed $\Theta/L =: 1$, in $y\rightarrow0$ limit
\begin{equation}\label{eq:AdSft_bdcond1}
1 = \frac{\Theta}{L} 
= \sqrt\frac{\cos^4\theta_0+\kappa^2}{(y'^2+x_3'^2+y^2\theta'^2)(1-\omega^2y^2\sin^2\theta_0) + g'(0)^2y^2\sin^2\theta_0}
\rightarrow
\sqrt\frac{\cos^4\theta_0+\kappa^2}{y'^2+x_3'^2}.
\end{equation}
In order to avoid the singularity at the boundary, the righthand side of the third equation \eqref{eq:timeintAdSfl_eom_3} is
\begin{equation}\label{eq:AdSft_bdcond2}
\frac{4y'}{y\Omega}(x_3'(1-\omega^2y^2\sin^2\theta) - \kappa)
\rightarrow
\frac{4y'}{y\Omega}(x_3'(0) - \kappa).
\end{equation}
Then, from eq.\eqref{eq:AdSft_bdcond1} and eq.\eqref{eq:AdSft_bdcond2} we impose the following condition:
\begin{equation}
y'(0) = \cos^2\theta_0,\;\;
x_3'(0) = \kappa.
\end{equation}
We impose the Neumann boundary condition for the angular direction $\theta'(0) = 0$.
The righthand side of the forth equation of motion \eqref{eq:timeintAdSfl_eom_4} approaches 
\begin{equation}
\frac{2g'}{y\sin\theta}(y'\sin\theta - y\theta'\cos\theta)
\rightarrow
\frac{2g'\cos\theta_0}{y\sin\theta_0}(\cos\theta_0\sin\theta_0 - y\theta'),\;\;
\therefore
g'(0) = 0.
\end{equation}
From the above discussion we obtain the appropriate boundary condition 
\begin{subequations}\label{eq:TiAdSfl_bdcond}
\begin{align}
&y(0) = 0,\;\;
\theta(0) = \theta_0,\;\;
x_3(0) = \kappa\sigma,\;\;
g(0) = 0,\\
&y'(0) = \cos^2\theta_0,\;\;
\theta'(0) = 0,\;\;
x_3'(0) = \kappa,\;\;
g'(0) = 0.
\end{align}
\end{subequations}

\subsection{Solution}
We solve the equations of motion \eqref{eq:TiAdSfl_4eom} under the initial condition \eqref{eq:TiAdSfl_bdcond}.
The results are summarized in the following seven figures.

\begin{figure}[h]
	\begin{minipage}[t]{0.5\linewidth}
	\includegraphics[width=\linewidth]{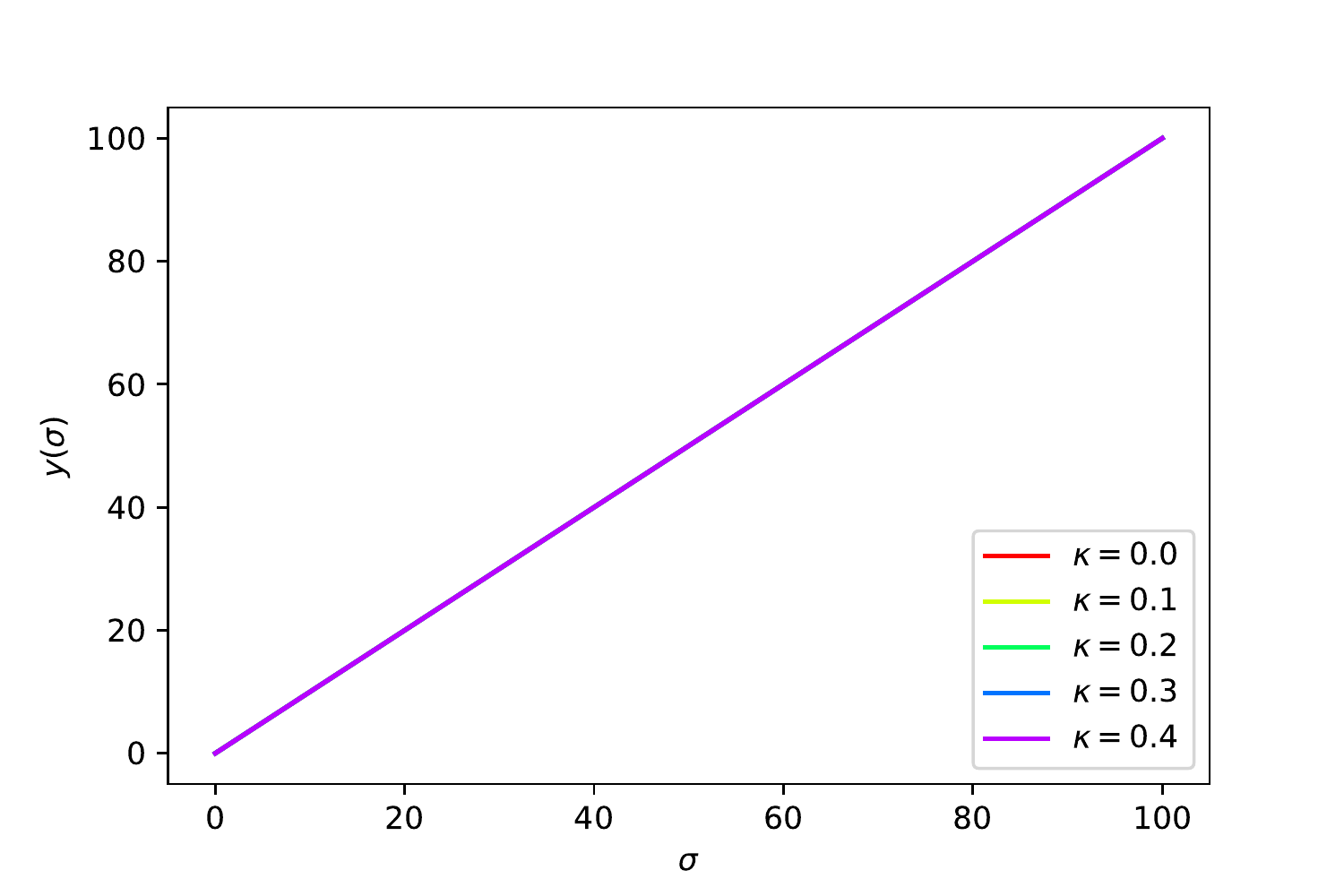}
	\caption{Increase of $y$ for $\theta_0=0$}
	\label{fig:iTAdS_k_y}
	\end{minipage}
\hspace{0\linewidth}
	\begin{minipage}[t]{0.5\linewidth}
	\includegraphics[width=\linewidth]{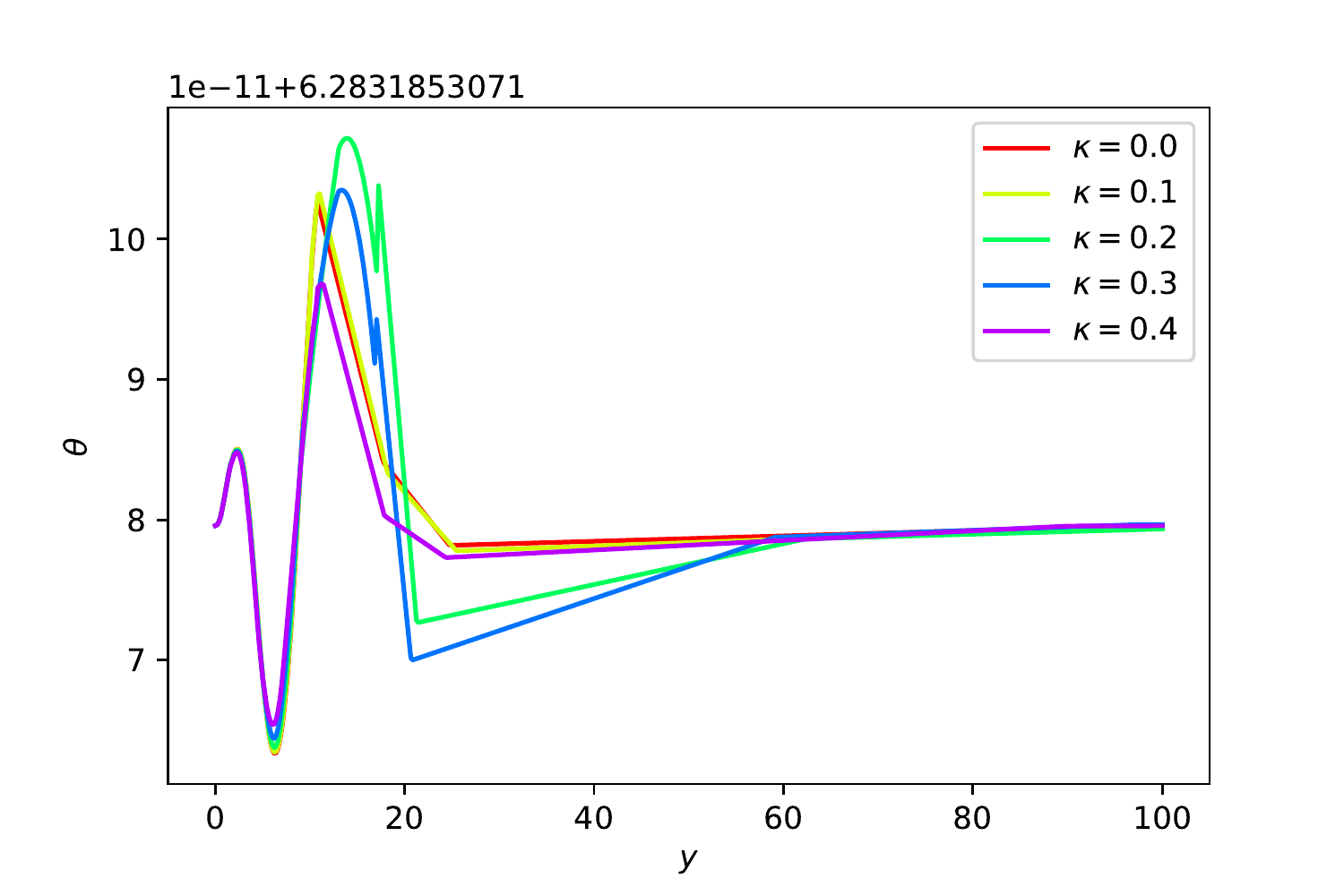}
	\caption{$\kappa$ dependence of $\theta$}
	\label{fig:iTAdS_K_theta}
	\end{minipage}
\hspace{0\linewidth}
	\begin{minipage}[t]{0.5\linewidth}
	\includegraphics[width=\linewidth]{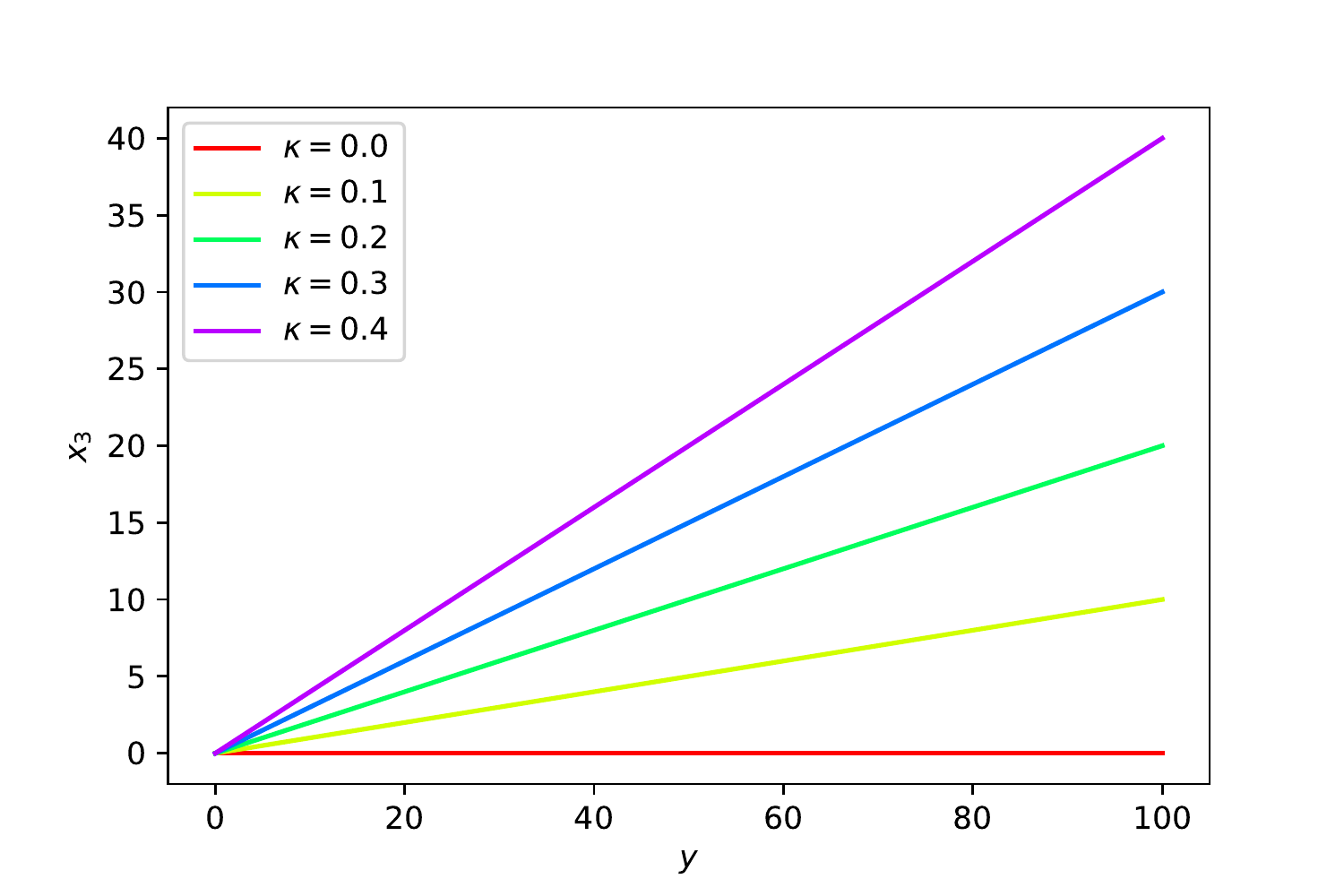}
	\caption{$\kappa$ dependence of $x_3$}
	\label{fig:iTAdS_K_x3}
	\end{minipage}
\hspace{0\linewidth}
	\begin{minipage}[t]{0.5\linewidth}
	\includegraphics[width=\linewidth]{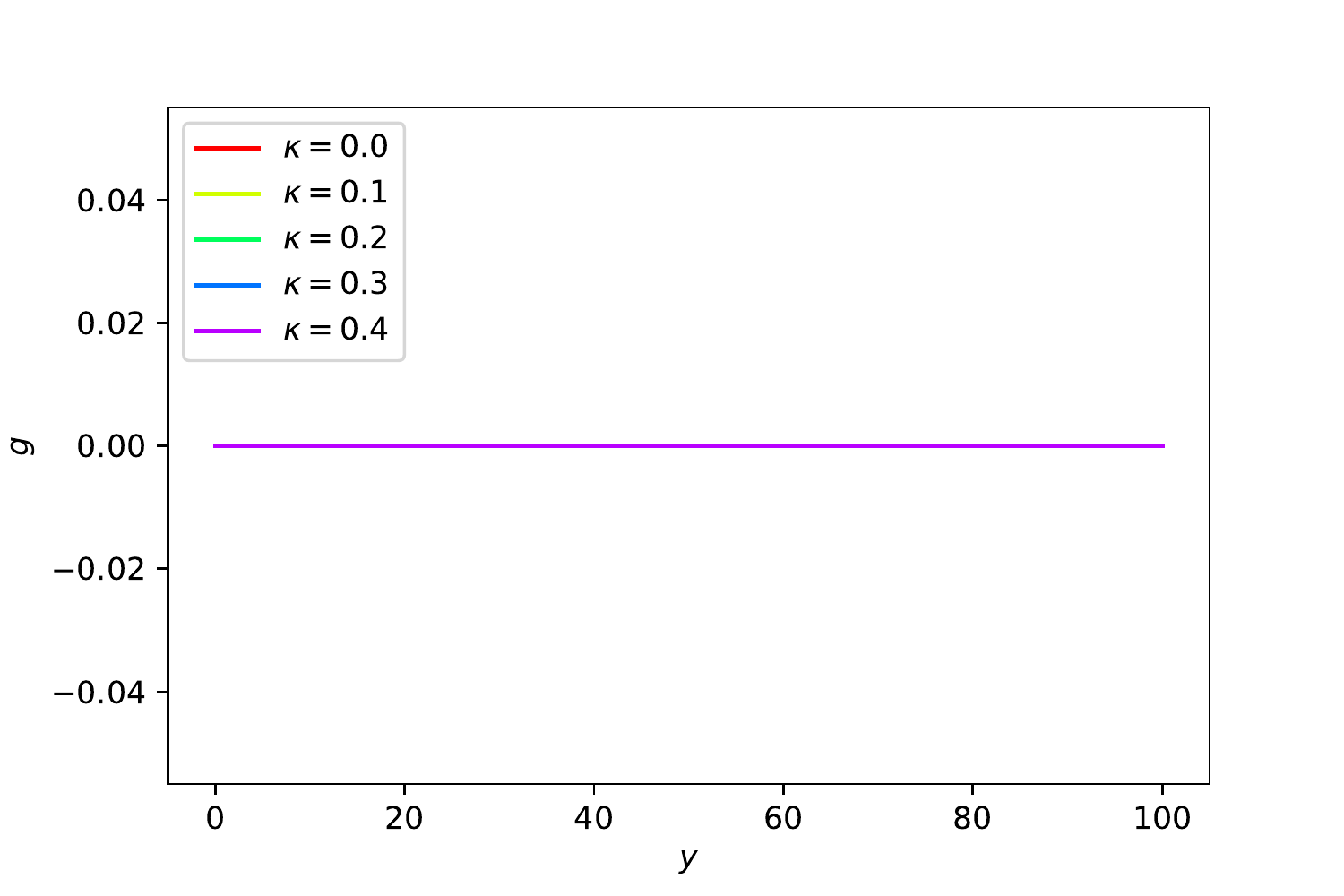}
	\caption{$\kappa$ dependence of $g$}
	\label{fig:iTAdS_K_g}
	\end{minipage}
\end{figure}

The function $y$ increases linearly for the value of $\sigma$ in the case $\theta_0 = 0$ as shown in Figure \ref{fig:iTAdS_k_y}.

The next three figures show the gauge flux dependence.
Figure \ref{fig:iTAdS_K_theta} and Figure \ref{fig:iTAdS_K_g} show the behavior of $\theta$ and $g$ as functions of $y$.
In these figures, neglecting numerical error, we read $\theta$ and $g$ do not grow from zero.
Figure \ref{fig:iTAdS_K_x3} shows linear relationship $x_3 = \kappa y$ as the same in \cite{Nagasaki:2011ue}.

\begin{figure}[h]
	\begin{minipage}[t]{0.5\linewidth}
	\includegraphics[width=\linewidth]{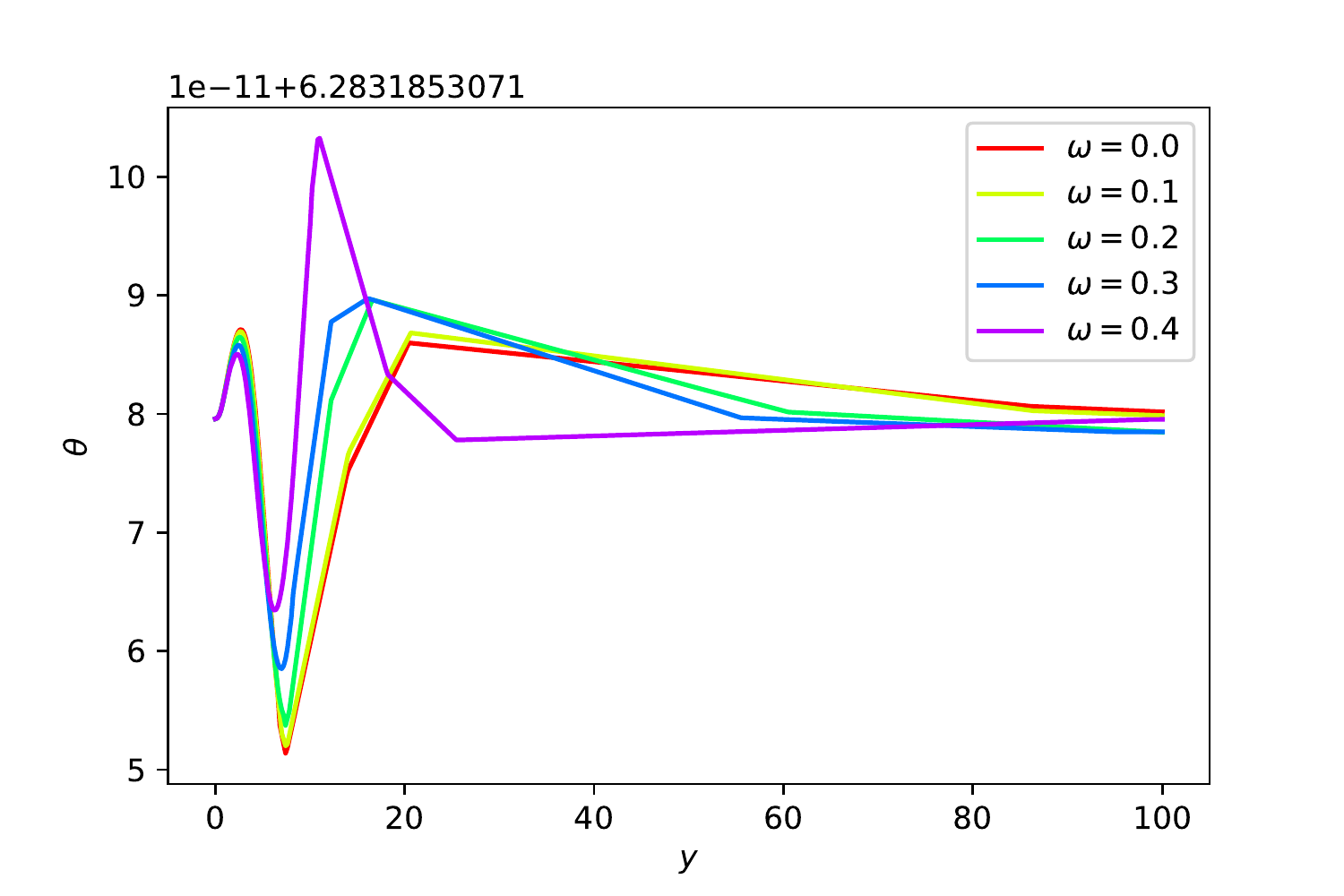}
	\caption{$\omega$ dependence of $\theta$}
	\label{fig:iTAdS_omega_theta}
	\end{minipage}
\hspace{0\linewidth}
	\begin{minipage}[t]{0.5\linewidth}
	\includegraphics[width=\linewidth]{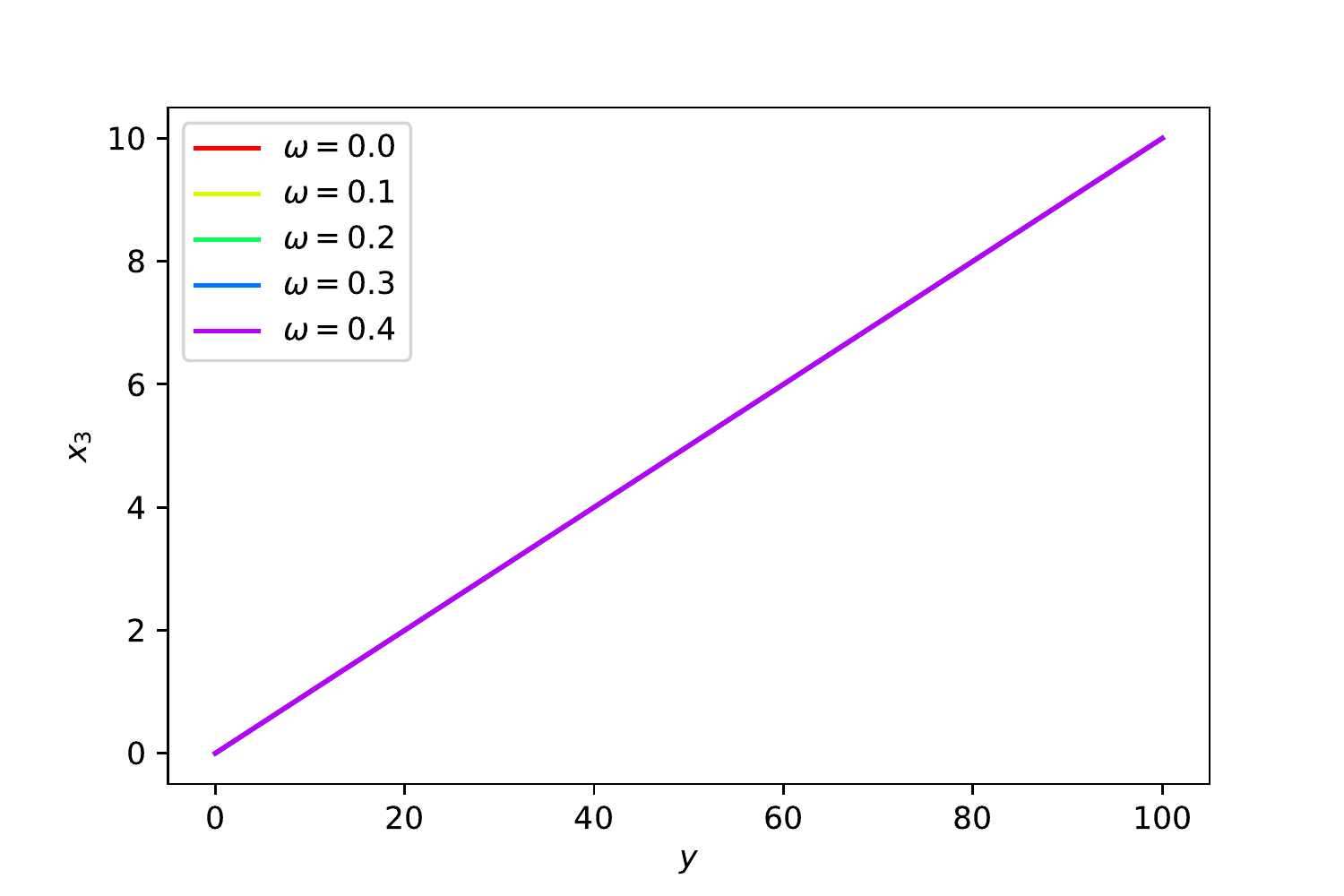}
	\caption{$\omega$ dependence of $x_3$}
	\label{fig:iTAdS_omega_x3}
	\end{minipage}
\hspace{0\linewidth}
	\begin{minipage}[t]{0.5\linewidth}
	\includegraphics[width=\linewidth]{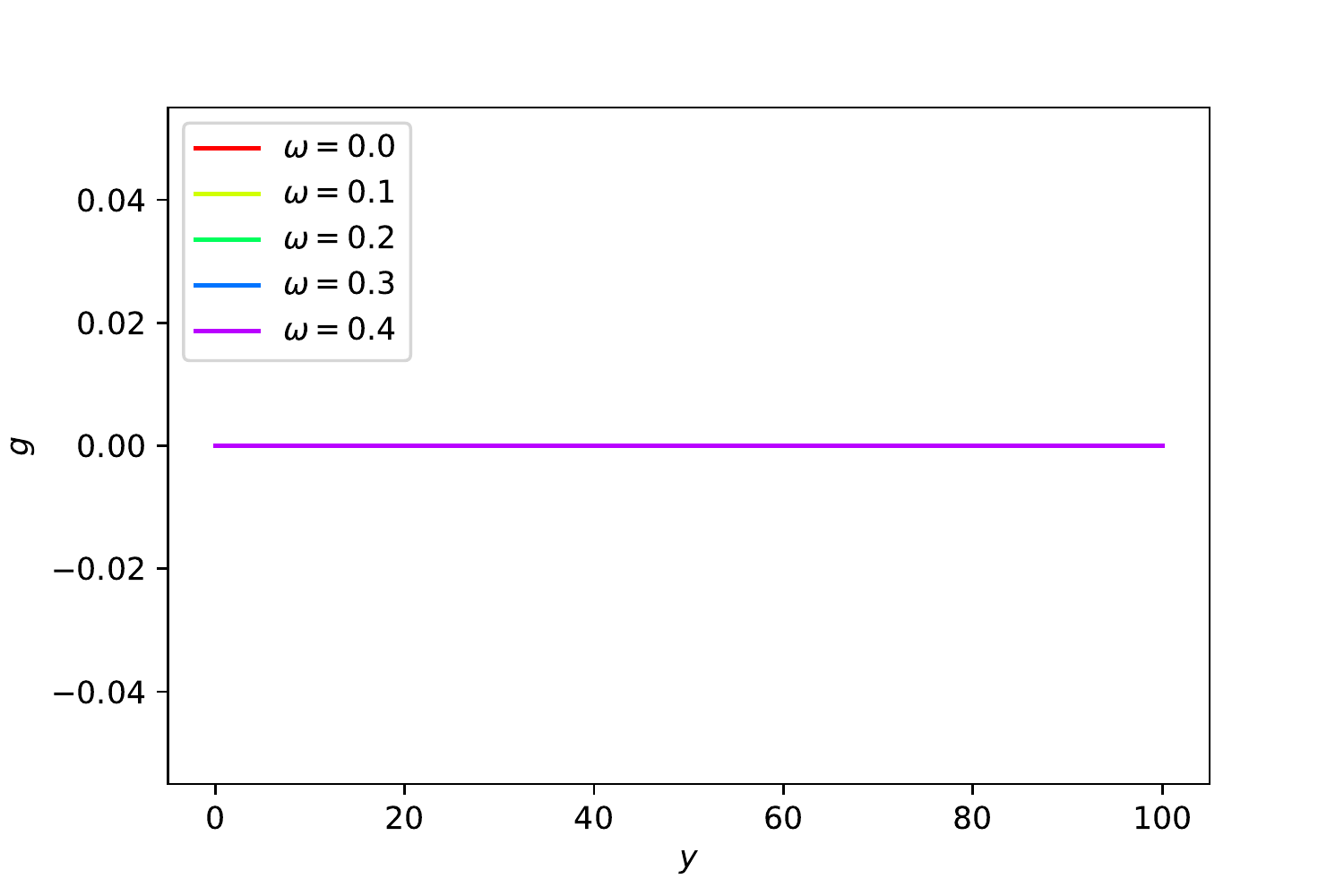}
	\caption{$\omega$ dependence of $g$}
	\label{fig:iTAdS_omega_g}
	\end{minipage}
\end{figure}

The next three figures show the angular momentum dependence (see Figure \ref{fig:iTAdS_omega_theta}, Figure \ref{fig:iTAdS_omega_x3} and Figure \ref{fig:iTAdS_omega_g}). 
From these results, we find that there is no dependence on the angular momentum.
It can also be seen from the $\theta$ plot (Figure \ref{fig:iTAdS_omega_theta}) and the fact that the $\omega$ dependence in eqs.\eqref{eq:TiAdS_4eom2} exists only in the combination with $\sin\theta$.

\section{Black hole spacetime}\label{sec:BHspacetime}
In this section we generalize the solution obtained in the previous section to the black hole spacetime.
The metric is
\begin{align}
ds_{\text{AdS}_5\times S^5}^2
&= \frac{r^2}{R_\text{AdS}^2}(-h dt^2 + d\vec x^2)
 + \frac{R_\text{AdS}^2}{r^2}\frac{dr^2}{h}
 + R_\text{AdS}^2ds_{S^5}^2,\\
h(r) &= 1-\frac{r_\text{H}^4}{r^4},
\end{align}
where $R_\text{AdS}$ is the radius of the AdS and the $S^5$ metric is, by coordinates 
$\theta,\psi\in[0,\pi/2]$ and $\varphi,\phi,\chi\in[0,2\pi)$,
\begin{equation}
ds_{S^5}^2
= d\theta^2 + \sin^2\theta d\varphi^2 + \cos^2\theta
  (d\psi^2 + \sin^2\psi d\phi^2 + \cos^2\psi d\chi^2).
\end{equation}
In the following, we put $R_\text{AdS}=1$.
By changing variables, $y = 1/r$, this metric becomes
\begin{equation}
ds_{\text{AdS}_5\times S^5}^2
= \frac1{y^2}(-h dt^2 + \frac{dy^2}{h} + d\vec x^2)
 + ds_{S^5}^2.
\end{equation}

In the same way as before, we consider the case where the D5-brane extends to the directions
\begin{equation}
t,x_1,x_2\in\text{AdS}_5,\;
\psi,\phi\in S^5.
\end{equation}
By parameter $\sigma$ the embedding of the D5-brane is 
\begin{equation}
y = y(\sigma),\;\;
x_3 = x_3(\sigma),\;\;
\theta = \theta(\sigma),\;\;
\varphi = \omega t + g(\sigma).
\end{equation}
The gauge flux is obtained in the same way as the flat case,
\begin{equation}\label{eq:TiAdSBH_gaugeflux}
\mathcal F = -\kappa\sin\psi d\psi\wedge d\phi.
\end{equation}
The RR4-from is 
\begin{equation}
C_4 = -\frac1{y^4}dtdx_1dx_2dx_3 + 4\alpha_4,
\end{equation}
where $\alpha_4$ is a 4-form satisfying $d\alpha_4 = \text{volume of }S^5$.
The induced metric becomes
\begin{align*}
ds_\text{D5}^2
&= -\Big(\frac{h}{y^2}-\omega^2\sin^2\theta\Big)dt^2
 + \Big(\frac{y'^2/h+x_3'^2}{y^2} + \theta'^2 + g'^2\sin^2\theta\Big)d\sigma^2
 + 2\omega g'\sin^2\theta dtd\sigma\nonumber\\
&\hspace{5cm}
 + \frac{dx_1^2 + dx_2^2}{y^2}
 + \cos^2\theta(d\psi^2 + \sin^2\psi d\phi^2).
\end{align*}
The sum of the induced metric and the gauge flux is 
\begin{align}
&G_\text{D5} + \mathcal F\nonumber\\
&=\begin{bmatrix}
-\frac1{y^2}(h-\omega^2y^2\sin^2\theta)& \omega g'\sin^2\theta& & & & \\
\omega g'\sin^2\theta& \frac1{y^2}(\frac{y'^2}h+x_3'^2+y^2\theta'^2)+g'^2\sin^2\theta& & & & \\
 & & 1/y^2& & & \\
 & & & 1/y^2& & \\
 & & & & \cos^2\theta& F'\\
 & & & & - F'& \cos^2\theta\sin^2\psi
\end{bmatrix}.
\end{align}
Then the DBI action is 
\begin{equation}
S_\text{DBI}/T_5
= -2\pi TV\int dyd\psi
  \frac{\sqrt{\cos^4\theta\sin^2\psi+F'^2}}{y^4}L(\sigma).
\end{equation}
The WZ term is
\begin{equation}
S_\text{WZ}/T_5 
= -2\pi TV\int d\psi dy\frac{F'(\psi)x_3'}{y^4}.
\end{equation}
Summing them the action is 
\begin{equation}
S_\text{D5}/T_5
= -2\pi TV\int\frac{d\psi dy}{y^4}
\Big[\sqrt{\cos^4\theta\sin^2\psi+F'^2}L(y) + x_3'F'\Big].
\end{equation}
We can determine the gauge flux in the same way as found in eq.\eqref{eq:TiAdSfl_gaugeflux2}.
Substituting the gauge flux \eqref{eq:TiAdSBH_gaugeflux}, the action is 
\begin{align}
S_\text{D5}/T_5
&= -2\pi TV\int\frac{dy}{y^4}
(\sqrt{\cos^4\theta+\kappa^2}L(\sigma) - \kappa x_3'),\\
L(\sigma) &= \sqrt{\Big(\frac{y'^2}h+x_3'^2+y^2\theta'^2\Big)(h-\omega^2y^2\sin^2\theta)+hy^2g'^2\sin^2\theta}.\nonumber
\end{align}

\subsection{Equations of motion}
For notational convenience, we define 
\begin{equation}
\Theta := \sqrt{\cos^4\theta + \kappa^2},\;\;
S := \frac{y'^2}{h}+x_3'^2+y^2\theta'^2,\;\;
\Omega := h - \omega^2y^2\sin^2\theta.
\end{equation}
In this notation the action is 
\begin{equation}
S_\text{D5}\sim \int d\sigma\frac{\Theta L - \kappa v}{y^4},\;
L = \sqrt{S\Omega+hg'^2y^2\sin^2\theta}.
\end{equation}
The equations of motion for $y$,$\theta$,$x_3$ and $g$ are
\begin{subequations}\label{eq:TiAdSBH_4eom}
\begin{align}
&\frac{d}{d\sigma}\Big(\frac{\Theta}{y^4}\frac{(y'/h)\Omega}{L}\Big)
 + \frac4{y^5}(\Theta L - \kappa v)
 - \frac{\Theta}{y^4L}(y\theta'^2\Omega+y(hg'^2-\omega S)\sin^2\theta)\nonumber\\
&\hspace{5cm}
 - \frac{\Theta}{y^4L}\Big(-\frac{y'^2\Omega}{h^2} 
   + S + y^2g'^2\sin^2\theta\Big)\partial h = 0,\\
&\frac{d}{d\sigma}\Big(\frac{\Theta}{y^4}\frac{\theta'y^2\Omega}{L}\Big) 
 + \frac{2\sin\theta\cos^3\theta}{\Theta}\frac{L}{y^4}
 - \frac{\Theta}{y^4}\frac{y^2(hu^2-\omega^2S)\sin\theta\cos\theta}{L} = 0,\\
&\frac{d}{d\sigma}\Big(\frac{\Theta}{y^4}\frac{x_3'\Omega}{L}
 - \frac{\kappa}{y^4}\Big) = 0,\\
&\frac{d}{d\sigma}\Big(\frac{\Theta}{y^4}\frac{hy^2\sin^2\theta}{L}g'\Big) = 0.
\end{align}
\end{subequations}
We choose the gauge $L/\Theta = 1$,
The equations of motion become
\begin{subequations}
\begin{align}
&\frac{d}{d\sigma}\Big(\frac{y'}{h}\frac{\Omega}{y^4}\Big)
 + \frac4{y^5}(\Theta^2 - \kappa v)
 - \frac1{y^3}(\theta'^2\Omega+(hg'^2-\omega S)\sin^2\theta)\nonumber\\
&\hspace{5cm}
 - \frac1{y^4}\Big(-\frac{y'^2\Omega}{h^2} 
   + S + y^2g'^2\sin^2\theta\Big)\partial h = 0,\\
&\frac{d}{d\sigma}\Big(\theta'\frac{\Omega}{y^2}\Big) 
 + \frac{2\sin\theta\cos^3\theta}{y^4}
 - \frac{(hu^2-\omega^2S)\sin\theta\cos\theta}{y^2} = 0,\\
&\frac{d}{d\sigma}\Big(\frac{x_3'\Omega}{y^4} - \frac{\kappa}{y^4}\Big) = 0,\\
&\frac{d}{d\sigma}\Big(\frac{h\sin^2\theta}{y^2}g'\Big) = 0.
\end{align}
\end{subequations}

We define the following:
\begin{subequations}
\begin{align}
A &:= -\frac{d}{d\sigma}\log\Omega
= \frac{2\omega^2y\sin\theta(y'\sin\theta + y\theta'\cos\theta) + 4r_\text{H}^4y^3y'}\Omega,\\
B &:= \frac{4h}{y\Omega}(\kappa v-\Theta^2),\;\;
C := \frac{hg'^2-\omega S}{\Omega},\;\;
D := -\frac{2\sin\theta\cos^3\theta}{y^2\Omega},\\
E &:= \frac{h(S + y^2g'^2\sin^2\theta)}{\Omega}\partial h.
\end{align}
\end{subequations}
By solving the equations for the second derivative terms,
\begin{subequations}\label{eq:TiAdSBH_4eom2}
\begin{align}
y'' &= \frac{4y'^2}y + hy\theta'^2 + y'A + B + Chy\sin^2\theta + E,\\
\theta'' &= \frac{2\theta'y'}y + \theta'A + D + C\sin\theta\cos\theta,\\
x_3'' &= \frac{4x_3'y'}y - \frac{4\kappa y'}{y\Omega} + x_3'A,\label{eq:timeintAdSHB_eom_3}\\
g'' &= - 2\theta'g'\cot\theta + \frac{2y'g'}y -\frac{y'g'\partial h}h.\label{eq:timeintAdSHB_eom_4}
\end{align}
\end{subequations}

\paragraph{Boundary condition}
The boundary condition is, since we fixed $\Theta/L=1$, in $y\rightarrow0$ limit,
\begin{equation}\label{eq:AdSBH_bdcond1}
\sqrt\frac{\cos^4\theta_0+\kappa^2}{(\frac{y'^2}h+x_3'^2+y^2\theta'^2)(h-\omega^2y^2\sin^2\theta)+hy^2g'^2\sin^2\theta}
= \sqrt\frac{\cos^4\theta_0+\kappa^2}{y'^2+x_3'^2} = 1.
\end{equation}
From the third equation of motion \eqref{eq:timeintAdSHB_eom_3}
\begin{equation}\label{eq:AdSBH_bdcond2}
\frac{4y'}{y\Omega}(x_3'(h-\omega^2y^2\sin^2\theta)-\kappa)
\approx\frac{4y'}{y\Omega}(x_3'-\kappa).
\end{equation}
From equations \eqref{eq:AdSBH_bdcond1} and \eqref{eq:AdSBH_bdcond2} we find the conditions,
\begin{equation}
x_3'(0) = \kappa,\;\;
y' = \cos^2\theta_0.
\end{equation}
We impose the Neumann condition $\theta'(0) = 0$ at the AdS boundary.
The forth equation \eqref{eq:timeintAdSHB_eom_4} gives
\begin{equation}
\frac{2g'}{y\sin\theta}(y'\sin\theta - y\theta'\cos\theta)
\approx\frac{2g'\cos\theta_0}{y\sin\theta}
  (\cos\theta_0\sin\theta_0 - y\theta'),\;
g'(0) = 0.
\end{equation}
Summarizing the appropriate boundary condition is
\begin{subequations}\label{eq:TiAdSBH_bdcond}
\begin{align}
&y(0) = 0,\;\;
\theta(0) = \theta_0,\;\;
x_3(0)\approx\kappa\sigma,\;\;
g(0) = 0,\\
&y'(0) = \cos^2\theta_0,\;\;
\theta'(0) = 0,\;\;
x_3'(0) = \kappa,\;\;
g'(0) = 0.
\end{align}
\end{subequations}

\subsection{Solution}
We solve the equations of motion \eqref{eq:TiAdSBH_4eom2} under the boundary condition \eqref{eq:TiAdSBH_bdcond}.
The results of the numerical calculation are shown in the next seven figures.

\begin{figure}[h]
	\begin{minipage}[t]{0.5\linewidth}
	\includegraphics[width=\linewidth]{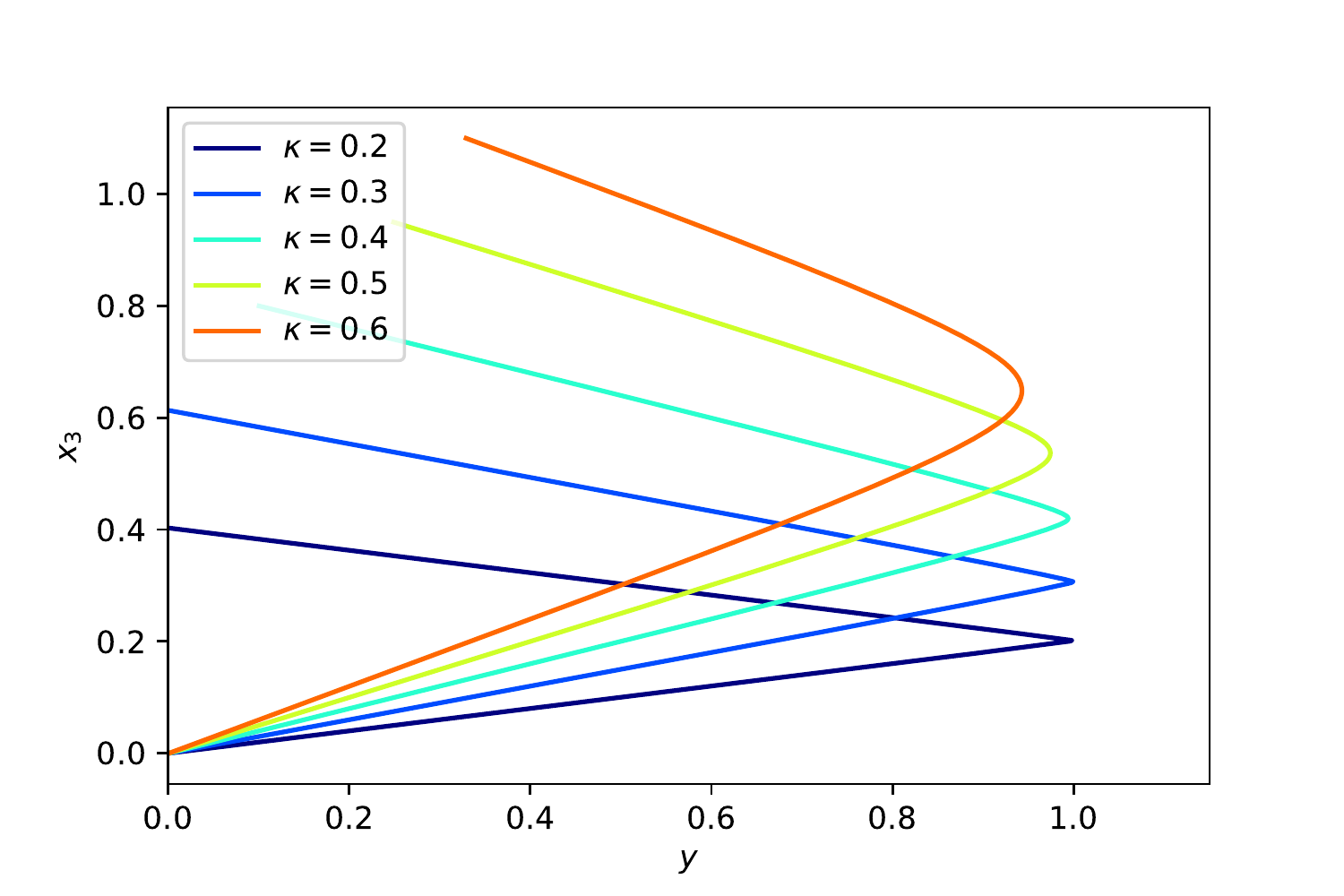}
	\caption{$\kappa$ dependence of $x_3$}
	\label{fig:iAdSBHx3_k_y}
	\end{minipage}
\hspace{0\linewidth}
	\begin{minipage}[t]{0.5\linewidth}
	\includegraphics[width=\linewidth]{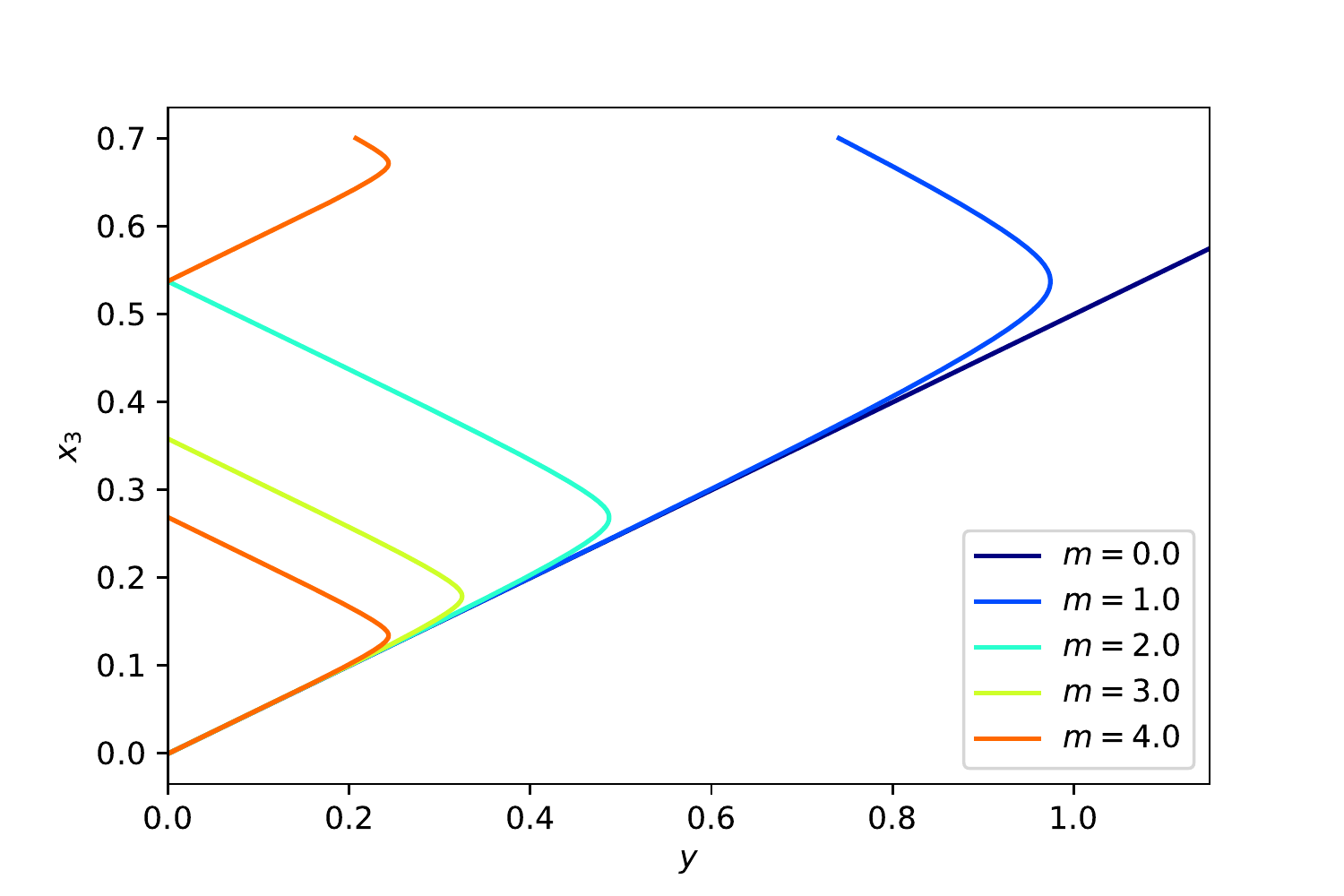}
	\caption{Mass dependence of $x_3$}
	\label{fig:iAdSBHx3_m_y}
	\end{minipage}
\hspace{0\linewidth}
	\begin{minipage}[t]{0.5\linewidth}
	\includegraphics[width=\linewidth]{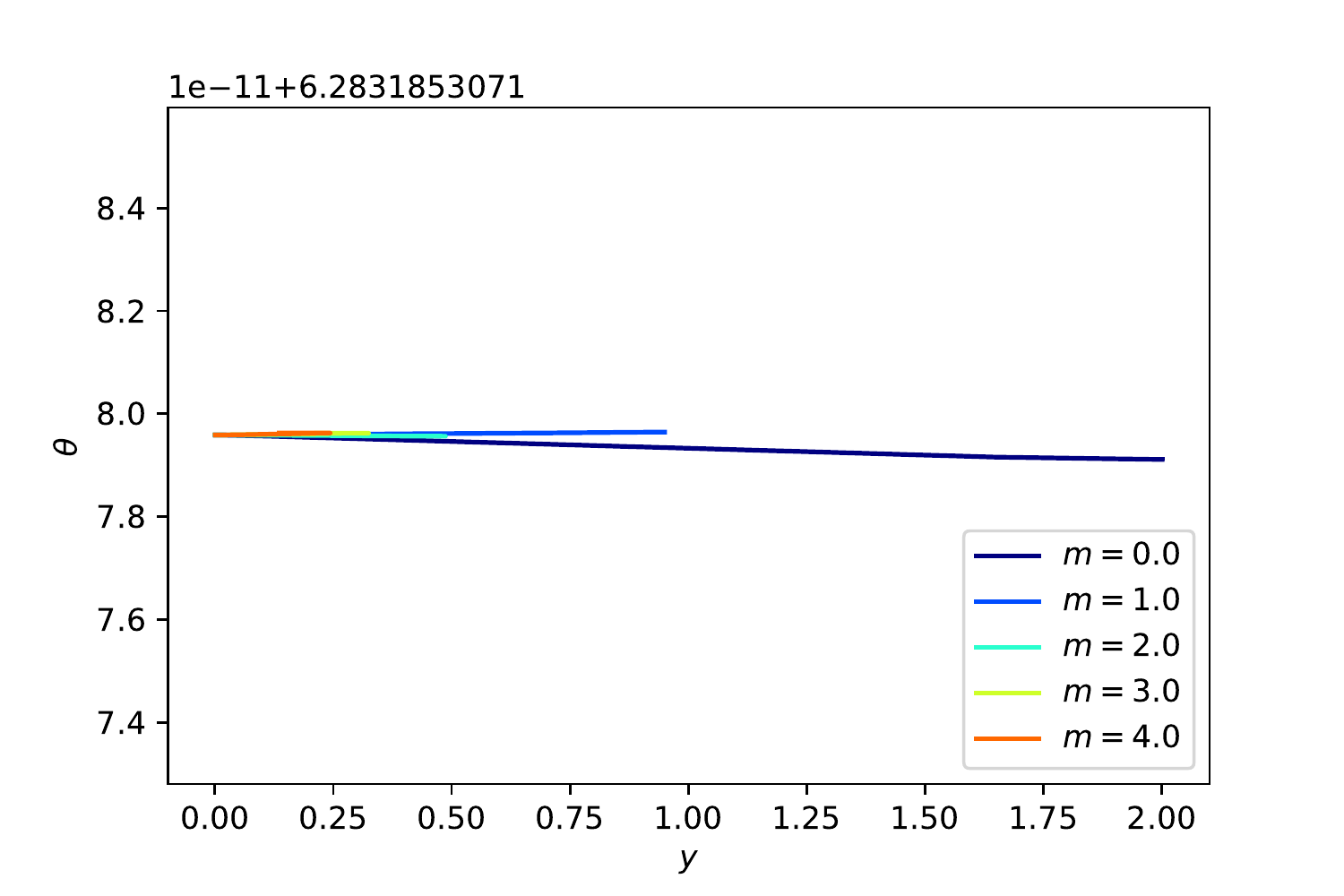}
	\caption{Mass dependence of $\theta$}
	\label{fig:iAdSBHx3_m_theta}
	\end{minipage}
\hspace{0\linewidth}
	\begin{minipage}[t]{0.5\linewidth}
	\includegraphics[width=\linewidth]{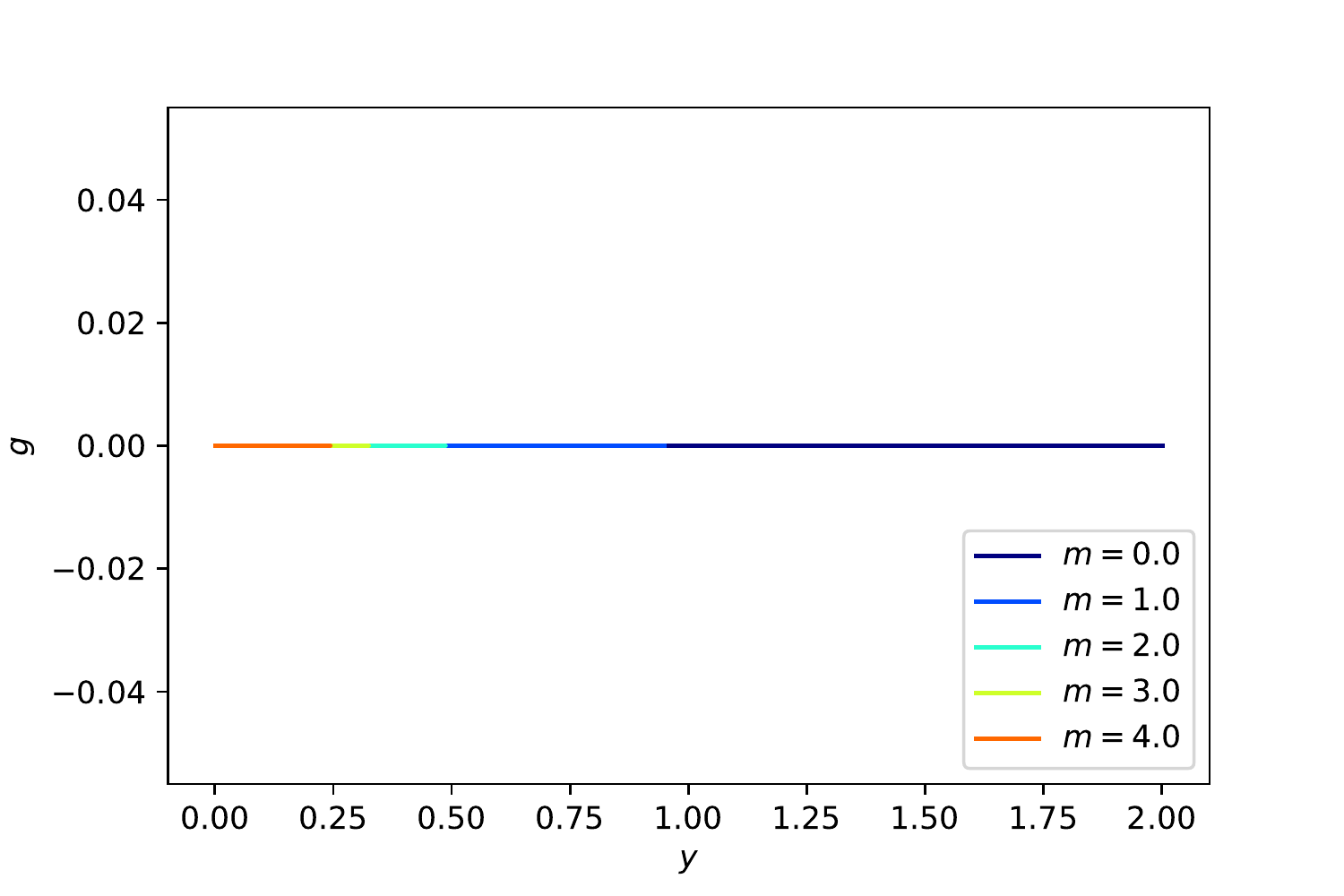}
	\caption{Mass dependence of $g$}
	\label{fig:iAdSBHx3_m_g}
	\end{minipage}
\end{figure}

Figure \ref{fig:iAdSBHx3_k_y} shows the behavior of $x_3$ as a function of $y$ for different gauge fluxes.
In this plot the angular momentum is $\omega = 0.5$ and the mass is $m=1$ as one can see the reflection point at the horizon $y=y_\text{h}$.
For large values of $\kappa$ the curve becomes smooth at the peak of $y$.

Figure \ref{fig:iAdSBHx3_m_y} shows the behavior of $x_3$ for different masses.
For $m=0$ the relation of $y$ and $x_3$ is linear.
For non-zero mass the curve bends in the vicinity of the horizon.
Since the equations of motion are not explicitly dependent on $x_3$, there are similar solutions which are parallel translated along $x_3$ direction.

As we can see in Figure \ref{fig:iAdSBHx3_m_theta} and Figure \ref{fig:iAdSBHx3_m_g}, $\theta$ and $g$ do not grow as the same for the previous section.

\begin{figure}[h]
	\begin{minipage}[t]{0.5\linewidth}
	\includegraphics[width=\linewidth]{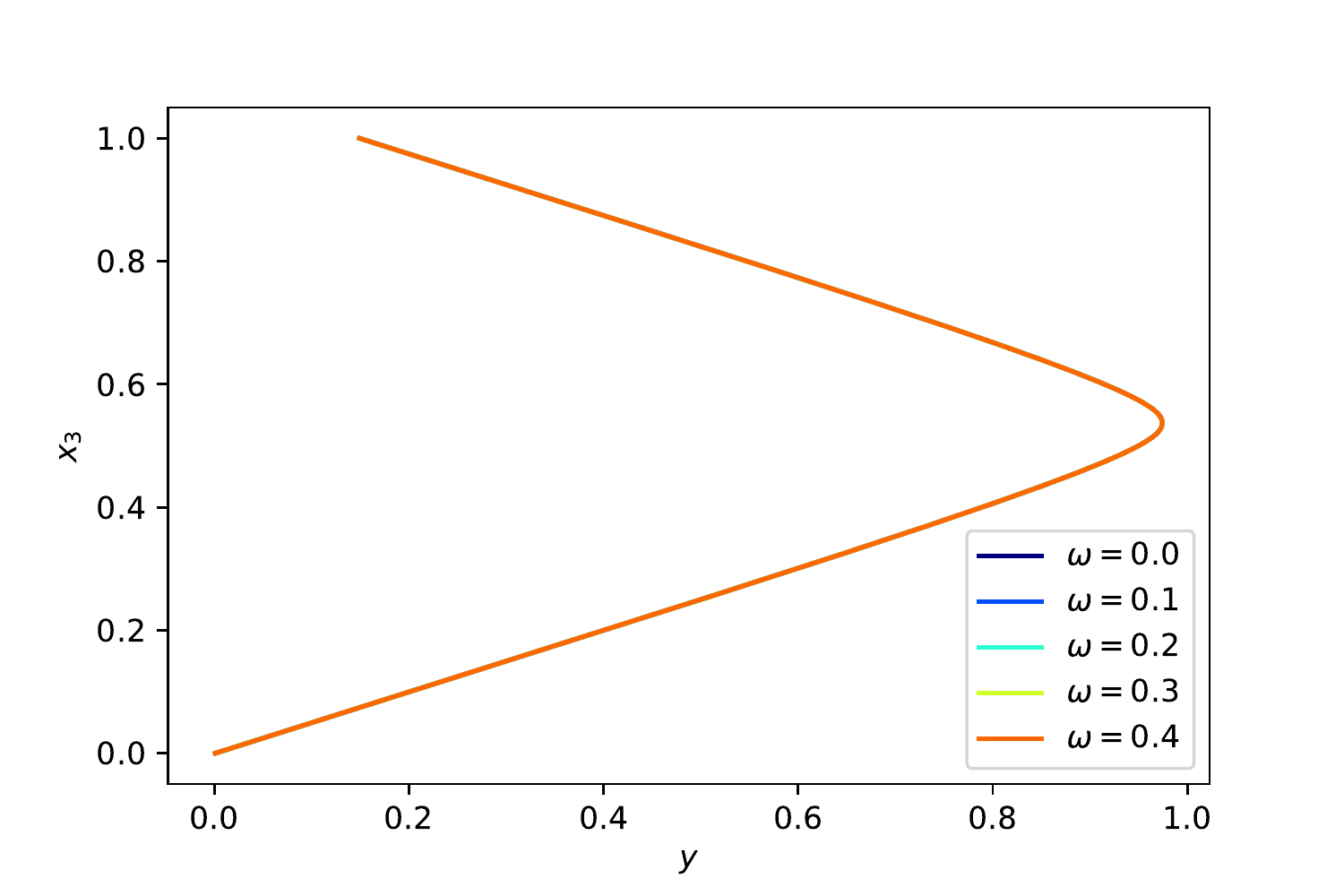}
	\caption{$\omega$ dependence of $x_3$}
	\label{fig:iAdSBHx3_omega_x3}
	\end{minipage}
\hspace{0\linewidth}
	\begin{minipage}[t]{0.5\linewidth}
	\includegraphics[width=\linewidth]{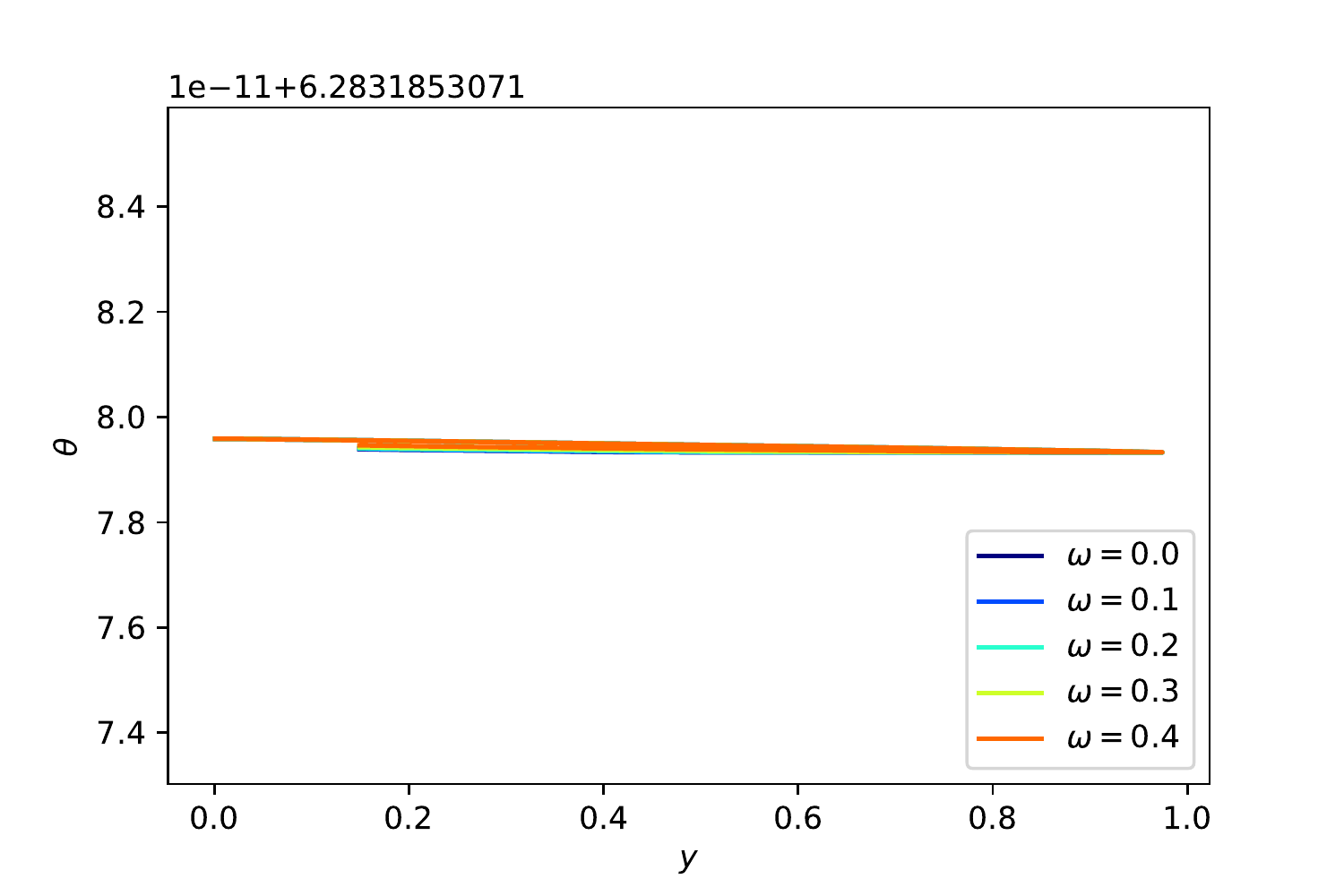}
	\caption{$\omega$ dependence of $\theta$}
	\label{fig:iAdSBHx3_omega_theta}
	\end{minipage}
\hspace{0\linewidth}
	\begin{minipage}[t]{0.5\linewidth}
	\includegraphics[width=\linewidth]{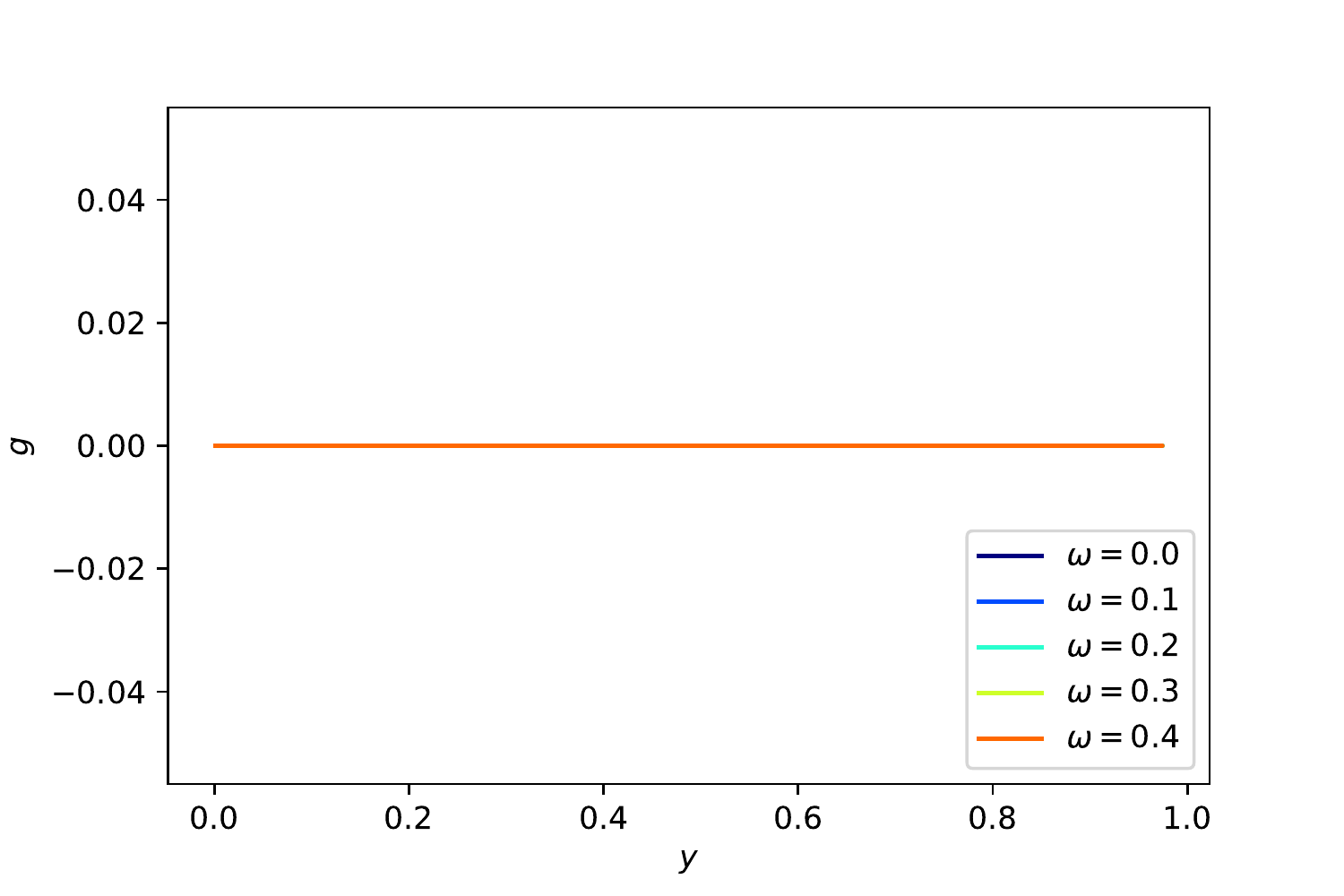}
	\caption{$\omega$ dependence of $g$}
	\label{fig:iAdSBHx3_omega_g}
	\end{minipage}
\hspace{0\linewidth}
	\begin{minipage}[t]{0.5\linewidth}
	\includegraphics[width=\linewidth]{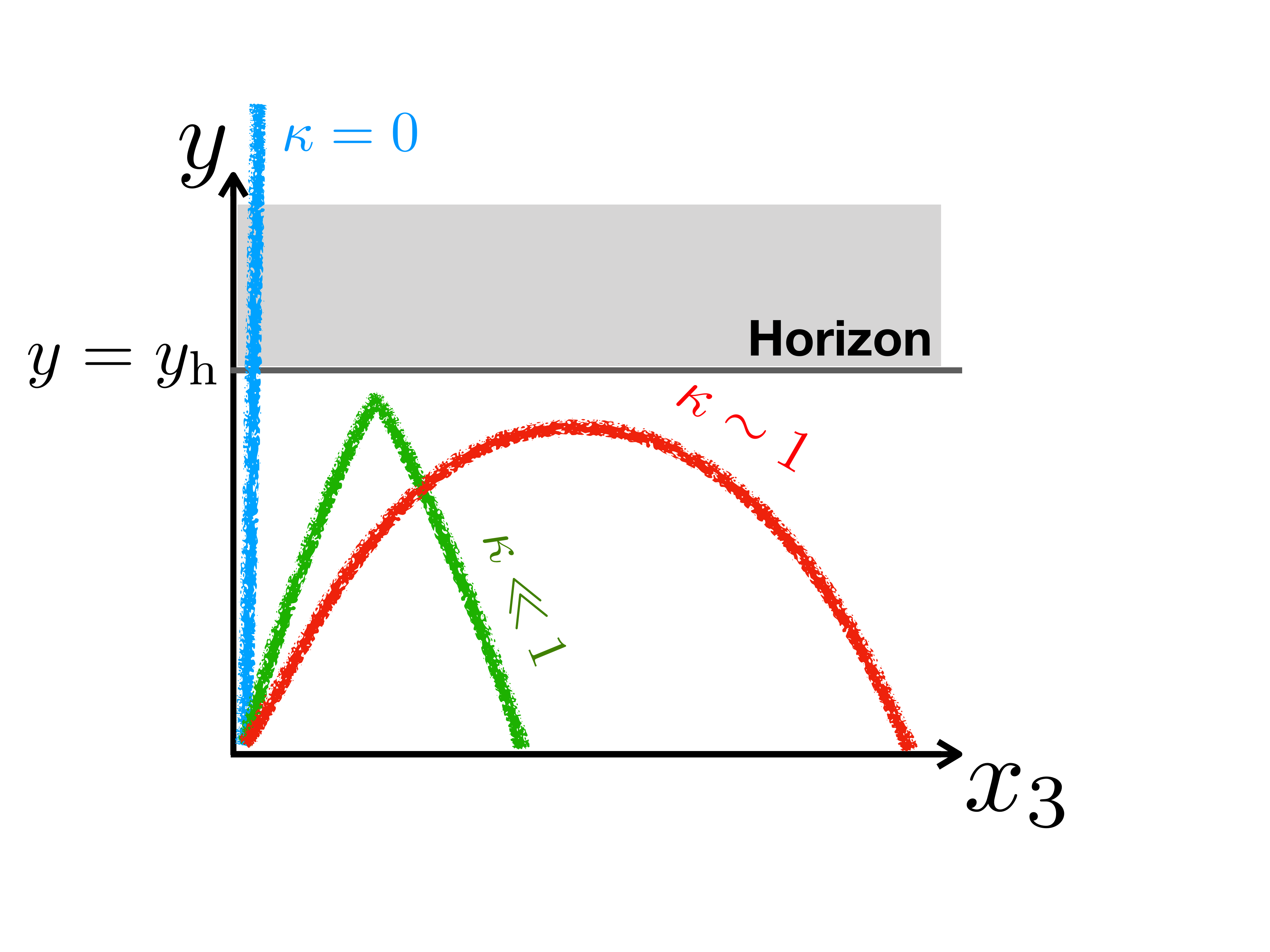}
	\caption{Form of the D5-brane}
	\label{fig:sketchik_x3_y}
	\end{minipage}
\end{figure}

Since $\theta=0$ is the solution, in the same way as before, there is no angular momentum dependence as shown in the next three figures (see Figure \ref{fig:iAdSBHx3_omega_x3}, Figure \ref{fig:iAdSBHx3_omega_theta} and Figure \ref{fig:iAdSBHx3_omega_g}).

In $\kappa=0$ case we choose the gauge $\sigma=y$.
Since identically $\theta=0$, each factor become $L = \sqrt{1+x_3'^2h}$, $\Theta = \sqrt{1+\kappa^2}$ and $\Omega = h$.
The equations \eqref{eq:TiAdSBH_4eom} are simplified as
\begin{equation}
\frac{d}{dy}\Big(x_3'\frac{h\sqrt{1+\kappa^2}}{y^4L}\Big)
= 0.
\end{equation}
We can see that $x_3'=0$ is the solution.
This is the unique solution which can penetrate the horizon. 
In this case the action integrated over the Wheeler DeWitt patch is simply, 
$
S_\text{D5}\sim\int_{y_\text{h}}^\infty dy/y^4
$.
The form of the D5-brane in ($x_3,y$)-plane is depicted in Figure \ref{fig:sketchik_x3_y}. 

\section{Discussion}\label{sec:Discussion}
In the plots of the solution we found the behavior of a moving interface in subspace $S^5$ of AdS$_5\times S^5$ black holes.
In this paper, we found two things:
The first is nontransparency of the horizon and the second is the independence of the angular momentum in $S^5$ subspace.
Let us discuss them in detail.

The D5-brane does not penetrate the horizon.
In complexity - volume \cite{Susskind:2014rva, Alishahiha:2015rta, Stanford:2014jda} and complexity - action \cite{Brown:2015lvg, Brown:2015bva} conjecture, we need to consider the surface or region which extends to the interior of the horizon.
Especially the growth rate of the action is calculated from the integration inside the horizon.
We want to find the effect of the non-local moving object such as \cite{Janiszewski:2011ue} to black hole complexity.
For this reason, we were looking for the case the probe brane penetrates the horizon.
In \cite{Nagasaki:2018dyz, Nagasaki:2019fln} we found that the static probe brane does not penetrate the horizon. 
Then we tried to find a time dependent solution in this paper.
According to the analysis, the probe D5-brane does not extend the horizon in the same way as the static case.
We restricted the motion of the D5-brane in $S^5$ subspace.
Then we found that the motion in $S^5$ part does not affect the embedding in AdS bulk.
In order to find the probe brane which affects CA conjecture, we may need to consider a system which has the motion in AdS bulk.
 
The second accomplishment is the discovery of the independence of the angular momentum in $S^5$.
That was due to the trivial solution for longitude coordinate $\theta=0$.
This result says the behavior of the probe brane in the AdS$_5$ part is decoupled from the motion in the $S^5$ subspace.

For this result, the interpretation in CFT side has another meaning because the action in AdS spacetime leads the drag force in the boundary theory \cite{Gubser:1998bc}.
In our case there are two interfaces on the boundary.
The gauge theory on the each side of the interface has gauge group SU($N$) and SU($N-k$).

In this work we focus on the motion restricted on a subspace of $S^5$. 
For future work, we are interested in time dependent solutions moving on the AdS subspace.
We expect to find a solution which can exist in the horizon for non-zero gauge flux.
This case gives non-trivial action which includes the gauge flux. 
This will be a generalization of (4.13) of \cite{Abad:2017cgl} which includes the nonzero gauge flux.
This gauge flux give us a useful way to compare the holographic quantities as stated in introduction.
Then, if we find the such time dependent solution, this give a good example to test  CA conjecture for AdS black holes. 

\section*{Acknnowledgments}
I would like to thank Sung-Soo Kim and Satoshi Yamaguchi for helpful discussion.

\providecommand{\href}[2]{#2}\begingroup\raggedright\endgroup

\end{document}